\documentclass{vgtc}                          

\ifpdf
  \pdfoutput=1\relax                   
  \usepackage{graphicx}                
  \DeclareGraphicsExtensions{.pdf,.png,.jpg,.jpeg} 
\else
  \usepackage{graphicx}                
  \DeclareGraphicsExtensions{.eps}     
\fi%

\graphicspath{{figures/}{pictures/}{images/}{./}} 

\usepackage{microtype}                 
\PassOptionsToPackage{warn}{textcomp}  
\usepackage{textcomp}                  
\usepackage{mathptmx}                  
\usepackage{times}                     
\usepackage{cite}                      
\usepackage{tabu}                      
\usepackage{booktabs}                  
\usepackage{orcidlink}
\usepackage{xcolor}

\onlineid{0}

\vgtccategory{Research}

\vgtcinsertpkg

\title{Experiences with CAMRE: Single-Device Collaborative Adaptive Mixed Reality Environment}

\author{Hung-Jui Guo\orcidlink{0000-0003-2233-846X}\thanks{Corresponding author: Hung-Jui Guo, e-mail: hxg190003@utdallas.edu}\\ %
        \scriptsize The University of Texas at Dallas %
    \and Omeed Eshaghi Ashtiani\orcidlink{0000-0002-6598-0551}\thanks{e-mail: omeed.ashtiani@utdallas.edu}\\ %
        \scriptsize The University of Texas at Dallas %
     \and Balakrishnan Prabhakaran\orcidlink{0000-0003-0385-8662}\thanks{e-mail: bprabhakaran@utdallas.edu}\\ %
     \scriptsize The University of Texas at Dallas %
     }

\teaser{
  \centering
  \includegraphics[width=0.8\textwidth, keepaspectratio=true]{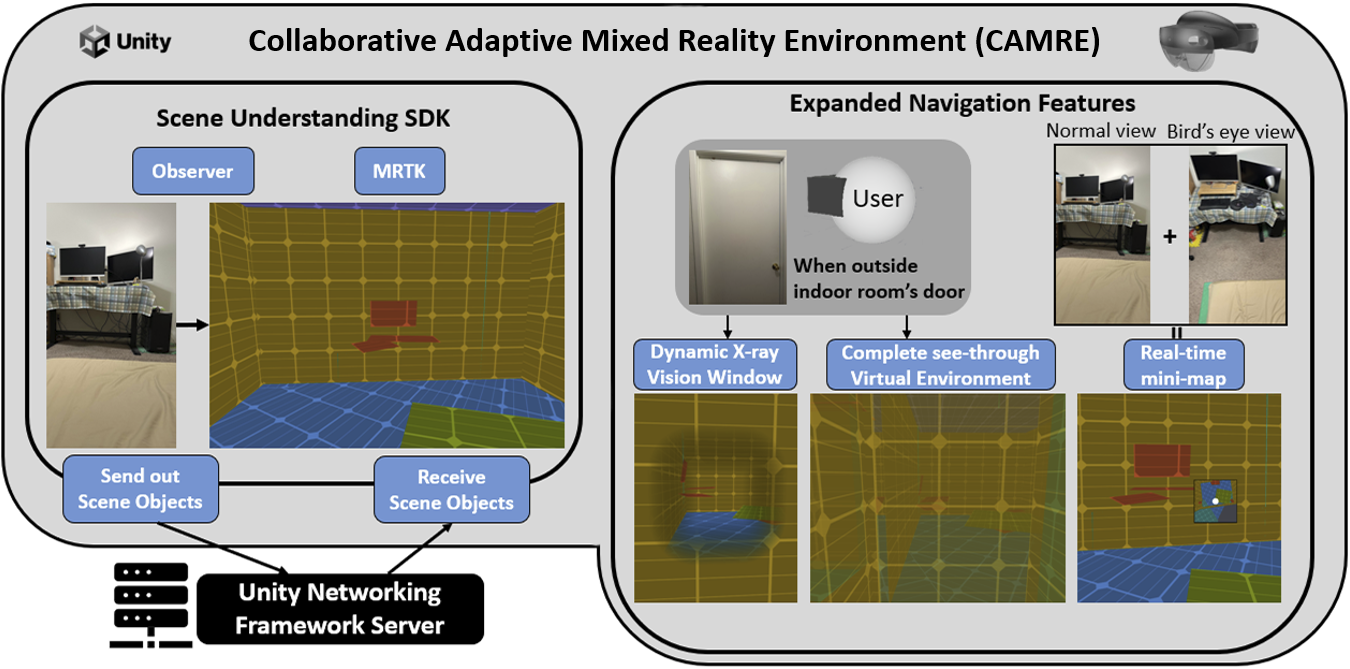}
  \caption{High-level architecture overview of the single-device CAMRE framework with external unity networking framework server.}
  \label{fig:Teaser}
}

\abstract{During collaboration in XR (eXtended Reality), users typically share and interact with virtual objects in a common, shared virtual environment. Specifically, collaboration among users in Mixed Reality (MR) requires knowing their position, movement, and understanding of the visual scene surrounding their physical environments. Otherwise, one user could move an important virtual object to a position blocked by the physical environment for others. However, even for a single physical environment, 3D reconstruction takes a long time and the produced 3D data is typically very large in size. Also, these large amounts of 3D data take a long time to be streamed to receivers making real-time updates on the rendered scene challenging. Furthermore, many collaboration systems in MR require multiple devices, which take up space and make setup difficult. To address these challenges, in this paper, we describe a single-device system called Collaborative Adaptive Mixed Reality Environment (CAMRE). We build CAMRE using the scene understanding capabilities of HoloLens 2 devices to create shared MR virtual environments for each connected user and demonstrate using a Leader-Follower(s) paradigm: faster reconstruction and scene update times due to smaller data. Consequently, multiple users can receive shared, synchronized, and close-to-real-time latency virtual scenes from a chosen Leader, based on their physical position and movement. We also illustrate other expanded features of CAMRE MR virtual environment such as navigation using a real-time virtual mini-map and X-ray vision for handling adaptive wall opacity. We share several experimental results that evaluate the performance of CAMRE in terms of the network latency in sharing virtual objects and other capabilities.
} 

\CCScatlist{
  \CCScatTwelve{Human-centered computing}{Mixed / augmented reality}{}{};
  \CCScatTwelve{Human-centered computing}{Collaborative interaction}{}{}
}

\begin{document}



\maketitle

\section{Introduction}
Multi-user collaboration in Virtual Reality (VR) and Mixed Reality (MR) has potential applications in a large variety of different fields, such as education \cite{lugrin2016breaking} and industrial settings \cite{pidel2020collaboration}. 
For physically distributed users, collaboration systems in Virtual Reality (VR) include networked, persistent, immersive, and virtual environments \cite{metaverse}.
Although this system is primarily for VR, the concept has been extended to MR. For example, in \cite{Piumsomboon2017}, the authors built a collaborative system combining Augmented Reality (AR) and VR devices to enable collaboration among users accessing different devices. By utilizing a Kinect camera, \cite{Fan2022} captured the user's motion and projected it onto a humanoid robot located in the collaborator's physical space to create an MR collaboration system. In the context of MR, users typically obtain views of their own surrounding physical environments while interacting with virtual objects. Therefore, unlike collaborative VR environments, MR collaboration systems will face several further issues.

\begin{figure}[htbp]
\centering
\includegraphics[width=0.45\textwidth, keepaspectratio=true]{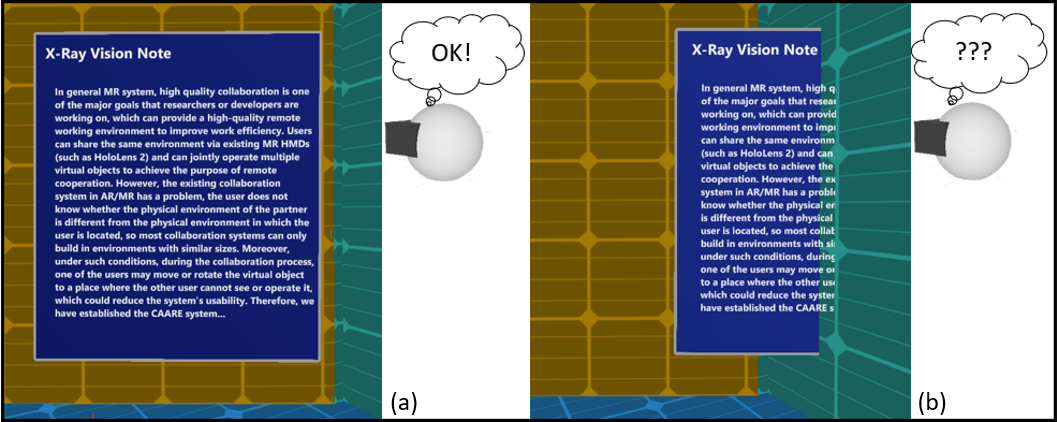}
\caption{(a) Example of one user may move or rotate the virtual object to a place or position where (b) the collaborator cannot see or operate it.}
\label{fig:Issue}
\vspace{-1em}
\end{figure}

Only a limited number of existing MR collaboration systems handle cases where some of the users' physical environments differ from the collaborators' current physical environments \cite{Wang2021}. Under such conditions, during the collaboration process, one of the users may move or rotate the virtual object to a place or position where the collaborator cannot see or operate it, which might have a negative effect on the collaboration process; an example is shown in Figure \ref{fig:Issue}.

\subsection{Challenges for Creating Physical Environment-based MR Collaboration System}
Creating a collaborative mixed reality system that employs a shared virtual environment based on the user's physical environment could lead to further challenges:

\begin{itemize}
    \vspace{-2mm}
    \item Computationally expensive to construct a 3-dimensional virtual environment based on the physical environment due to large data size; for instance, it would take about 100 Megabytes for a complete 3D indoor environment.
    \vspace{-2mm}
    \item Constructing a shared virtual environment to build a collaboration system often requires setting up multiple devices, which could be difficult for users unfamiliar with MR to get started and use the collaboration system.
    \vspace{-2mm}
    \item Transferring large-scale 3D environments can result in large streaming and update latencies over the Internet.
    \vspace{-2mm}
    \item Users tend to only use the virtual contents within their line of sight for collaboration and may have a limited understanding of the entire environment in MR virtual environments. This limited understanding of the overall environment might restrict physical movements and constrain the usage of the collaboration system.
    \vspace{-2mm}
\end{itemize}


\subsection{Collaborative Adaptive Mixed Reality Environment (CAMRE)}
To address the above challenges, we established a Collaborative Adaptive Mixed Reality Environment (CAMRE) system with one single MR device - Microsoft Hololens 2 \cite{hololensWeb} to not only share virtual objects but also share the virtual environment established based on one user's physical environment and share with other users that are connected to the same server, see Figure \ref{fig:Teaser} for an overview. 
To address the first challenge of using large 3D environments, instead of constructing the whole physical environment into mesh data, we use the scene understanding feature in the Microsoft Mixed Reality Toolkit (MRTK) \cite{hololensMRTK} to build virtual objects and a virtual environment based on the objects and geometry of the physical environment, which reduces data size from around 100 Megabytes (a living room size 3D mesh data) to approximately 0.3 Megabytes (a living room size scene understanding-based virtual environment). 
By utilizing the scene understanding feature to create a virtual environment, users are only required to deploy the CAMRE system on their HoloLens 2 device instead of setting up multiple sensors in the environment where they are located, which tackles the second challenge in terms of ease of use. 

Based on the created small-sized virtual environment, when users physically move in their physical environment, CAMRE can update the virtual environment accordingly by transmitting a small amount of data. On the basis of this system structure, we incorporated a \textbf{Leader-Follower paradigm}; the Leader is responsible for observing and creating an MR virtual environment and sharing with multiple Followers through a networking framework. CAMRE helps Leader and Followers share the same knowledge of the virtual environment, thereby preventing users from moving virtual objects to places where other collaborators cannot see. With CAMRE, the Leader can stream virtual information to Followers with minimal data usage, addressing the third challenge listed above.

\subsection{CAMRE's Expanded Navigation Features}
CAMRE with its Leader-Follower paradigm provides a method for one Leader and multiple Followers to collaborate in a shared virtual environment. However, since the occlusion of the wall objects generated from the Leader's side, users might stay in the initial room due to a limited understanding of other rooms, which will lead to the fourth challenge. 
Therefore, to tackle this challenge, we incorporated three expanded navigation features into the CAMRE system to provide an overview of the created virtual environment before physically moving to the destination to assist navigation and increase usability. 
\begin{itemize}
    \vspace{-2mm}
    \item \textbf{Dynamic X-ray vision}: Allows users to see through the surrounding obstacles to gain additional information about another room. (This feature was published separately in our demo paper; to ensure anonymity in reviewing process, we marked the authors' names as A. Anonymous in \cite{guo2022dynamic}.)
    \vspace{-2mm}
    \item \textbf{Complete see-through virtual environment}: Virtual walls will become partially transparent whenever the user approaches within 3 meters to provide information about all other rooms within range. 
    \vspace{-2mm}
    \item \textbf{Real-time mini-map}: Leader can observe and build the entire CAMRE MR virtual environment to share with Followers. Followers can explore either with the Leader or {\it independently} without following the Leader. This is facilitated by the real-time mini-map that shows the bird's eye view of the whole virtual environment and provides a complete view separately for each user. This mini-map feature makes an explicit assumption that such a complete view is available before starting the collaboration process and is given to the Followers.
    \vspace{-2mm}
\end{itemize}

When users are immersed in a virtual environment, their movements and interactions are significantly influenced by human depth perception \cite{Luo2007, Jones2008}. Unlike real objects with fixed images such as size and color in the human brain, users in virtual environments frequently lack adequate references to make accurate judgments regarding the depth perception of virtual objects due to the Vergence-accommodation conflict. Therefore, providing users with additional depth information in the virtual environment could help users have a better understanding of the surroundings. For example, \cite{Jones2011} presented a series of virtual environment underestimation experiments to suggest that visual information is an important source of information for the calibration of movement. 
In CAMRE, besides providing an overview of the virtual environment, the three expanded features could also provide additional depth information to assist users with navigation further. Dynamic X-ray vision can provide motion parallax since the X-ray vision window is moving with the user's eye gaze direction. A complete see-through virtual environment can provide distance perception due to the 3-meter setting that makes virtual walls partially transparent when users move within 3 meters of distance. Real-time mini-map provides camera position and field-of-view (FOV) options for users to adjust to provide relative distance and scale of the virtual objects. Here, we make an explicit assumption that the needed information such as the scene behind the obstacles is available (perhaps, through a pre-captured database) to the user. Detailed information will be provided in section \ref{CAMRE_Expanded}.

\subsection{Contributions}
We designed an exhaustive set of experiments with a primary objective of measuring latencies incurred during a Leader-multiple Followers collaboration over the Internet involving different distances among the collaborators. These experiments were conducted with varying factors such as room sizes, networking frameworks for sharing virtual environments, different distances between the Leader and Followers, and the number of simultaneous network connections. We make the following contributions through the created CAMRE framework:
\begin{enumerate}
    \vspace{-2mm}
    \item Implemented a user-friendly, single-device setup system for users who are unfamiliar with the MR system.
    \vspace{-2mm}
    \item Dynamically update virtual environment based on the corresponding physical environment with low data scale and low building time.
    \vspace{-2mm}
    \item Low streaming latency sharing virtual environment from Leader to multiple Followers to achieve Leader-Follower paradigm.
    \vspace{-2mm}
    \item Provide expanded navigation features to help users gain an overview of the created virtual environment to provide additional scene information and depth information.
    \vspace{-2mm}
    \item The extensive performance evaluation carried out on CAMRE involves two commonly used Unity networking frameworks and the performance results can serve as a benchmark network for other similar, future systems.
    \vspace{-2mm}
\end{enumerate}
Although some previous works created collaborative systems across multiple AR/VR/MR systems, to the best of our knowledge, the CAMRE system may be one of the earliest MR collaborative systems that include dynamic environmental updates and real-time ability.

\subsection{Using CAMRE}
We will make the CAMRE system software to be available as open source (after the paper is published). The CAMRE system along with the planned future work described in Section \ref{section-conclusion} can be very useful for the research community as well as application developers dealing with collaborative use cases such as training and tele-mentoring using MR. We will also make the experimental data reported in Section \ref{section-results} publicly available. The research community can use this data as a benchmark for comparing similar approaches. The data pertaining to network latencies in Section \ref{section-results} can also be used for trace-driven simulation for human subject studies in Internet-based collaborative MR applications.

\subsection{Limitations of Our Work}\label{section-limitations}
We also acknowledge some important limitations:
\begin{enumerate}
    \vspace{-2mm}
    \item Some previous MR collaboration systems (reported above) handle cases where some of the users’ physical environments differ from the collaborators’ current physical environments. However, CAMRE specifically employs a single Leader-multiple Followers paradigm, resulting in a common, shared virtual environment for all the users. While this could be a limitation for some use cases, the shared common virtual environment could be advantageous for training or telementoring types of applications.
    \vspace{-2mm}
    \item The performance studies reported in Sections \ref{section-experimental-design} and \ref{section-results} have been focused on network latencies in collaboration over the Internet. We have not carried out human subject studies to understand their perception of the effect of {\it degraded} (or small-data sized) virtual environments, nor on the effect of varying Internet latencies. 
    \vspace{-2mm}
    \item In a similar manner, our work has not evaluated the human perception of the effect of such a {\it synchronized} virtual environment as that of CAMRE. For instance, when the Leader moves to a different environment, the Followers' view/understanding of the virtual environment would also change accordingly even though they (the Followers) never move. This unexpected change in the environment might affect user experience and/or cause VR sickness.
    \vspace{-2mm}
\end{enumerate}

The above aspects of human perceptions need to be evaluated thoroughly. Considering the need for detailed and exhaustive user perception studies, we plan to do this as a future, separate research work. As mentioned earlier, will use the network latencies reported in Section \ref{section-results} to emulate Internet-level collaboration for these human perception studies.


\section{Related Work}
Many studies have been conducted to develop multi-user collaboration systems in AR/VR/MR that enable effective remote collaboration, particularly during the COVID-19 pandemic. One of the most common types of collaboration systems involves creating a virtual environment where users can immerse themselves and interact with other users' avatars to achieve collaborative outcomes. 
The concept of this type of system was proposed and discussed in 1998 \cite{Benford1998} as "collaborative virtual environments," which used networked virtual reality systems to support group work. 
More recently, various techniques have been used to achieve collaborative virtual environments; for example, \cite{Teo2019, Teo2020} presented a 360Drops system to provide 360 video sharing and 3D reconstructed scenes with photo-bubble to provide environment details. 
Additionally, researchers have been working on developing cross-reality systems to enable multiplayer collaboration across various AR and VR devices \cite{Tanaya2017, Piumsomboon2017}. 
\cite{Zhang2023} developed a VRGit system to facilitate multi-user collaboration in VR, which helps users manage the different versions of 3D content in virtual environments, making it easier for them to collaborate effectively. With this system, users can easily keep track of the modifications made to the virtual environment and manage the different versions of the content.
To conclude the development of collaborative virtual environments, a comprehensive survey on collaboration and communication systems was conducted by \cite{Druta2021} and \cite{Schafer2022} to provide insights into the different functionalities, advantages, and disadvantages of each collaboration system.

However, existing works rarely focused on sharing the whole surrounding physical environment, which could lead to the occlusion issue addressed in Figure \ref{fig:Issue} and reduce collaboration efficiency and freedom of moving in the virtual environment.
Still, some previous works have tried to share the surrounding physical environment to achieve collaboration; for instance, \cite{Duval2014} proposed a model to include users' surrounding physical environment into the virtual environment to build a collaborative environment in the VR world by taking into account the physical features and embedding them in the virtual environment.
The authors of \cite{Nittala2015} introduced a system called PLANWELL that utilized handheld AR devices for scanning outdoor geographical data by an explorer to create a 3D model, which could be shared with an overseer for remote collaboration. Although this system is similar to our Leader-Follower paradigm, it has a higher data transfer time (2.4 seconds), which might not be suitable for real-time collaboration.
\cite{Wang2021DistanciAR} presented a DistanciAR system that captured and created a remote environment with a LiDAR camera for viewing from a different location and improved the interface by adding Dollhouse (bird's-eye view) and Peek modes. However, the time spent using the complete system took around 13 minutes, which may pose a challenge when used for collaboration purposes. 
Most recently, \cite{Tian2023} presented a 3D MR remote collaboration prototype system by scanning the surrounding physical environment. However, to achieve real-time collaboration, this system utilized AR and VR head-mounted devices with three depth cameras to build 3D models by utilizing pre-scanned reconstructed 3D mesh models of a room-scale workspace, which could be challenging for regular users to set up.

In addition, other research works have tried to use humanoid robots to accomplish multi-player collaboration to perform actions captured by users that could potentially solve the occlusion issue. 
For example, \cite{Nagendran2015} used humanoid robots to imitate users' activities as surrogates to achieve cross-country collaboration, and \cite{Oyekoya2013} proposed a system integrating humanoid robots and video streams to build an MR-like collaborative environment for remote collaboration. 
To achieve better human-robot collaboration, \cite{Fan2022} suggested that robots should be capable of perceiving and parsing a scene’s information in real-time. The authors claimed that such environmental parsing is typically divided into three categories: Scene graph, 2D map representation, and 3D map representation, which echoes back to our scene understanding-based CAMRE system that performs scene understanding to build scene graphs with 3D map representation by reconstructing a corresponding 3D model from the understood information, and real-time mini-map to achieve 2D map representation.

The above summary of past and recent multi-user virtual environment works demonstrates a focus on sharing the same virtual environment in AR/VR. We believe that utilizing MR to immerse individuals in a common virtual environment based on their specific physical surroundings can provide additional information to facilitate collaboration.
Although similar collaborative systems and expanded features have been presented in other works built in AR/VR, our CAMRE system is one of the few collaborative systems built in MR, enabling users to interact with both the virtual environment and their physical environment. 
Furthermore, CAMRE utilizes a low data scale virtual environment to achieve low data transfer latency while still providing real-time collaboration with expanded features to enhance user experience and provide user-friendliness by providing accessible setup attributes with a single device.
In the following sections, we will focus on the detailed settings and evaluations of the CAMRE MR virtual environment with expanded low-latency sharing and expanded navigation features based on the surrounding physical environment.

\section{CAMRE System Design}
As mentioned earlier, in CAMRE, we employ a Leader-Follower paradigm through which MR environments are adaptively generated with low latency based on the physical environment of the Leader and shared using the open-source network frameworks in Unity (Unity Netcode for Gameobject \cite{NetcodeWeb} with Unity Relay \cite{RelayWeb} and Photon Unity Networking \cite{PhotonWeb}) for collaboration among multiple users. 
Instead of making every user hold the same level of authority, the Leader-Follower paradigm is employed to avoid multiple users building and sharing their virtual environment to cause virtual environment overlays, and only the Leader is authorized to observe and create the virtual environment.
Next, we add three expanded features (dynamic X-ray vision window, complete see-through virtual environment, and a real-time mini-map; see below) to help users navigate and gain depth cues of the adaptive MR environments. This system is designed and built on the Microsoft HoloLens 2 and HoloLens Unity emulator.


\subsection{MR Virtual Environment Adapting to Physical Environment}
\begin{figure*}[t]
\centering
\includegraphics[width=0.9\textwidth, keepaspectratio=true]{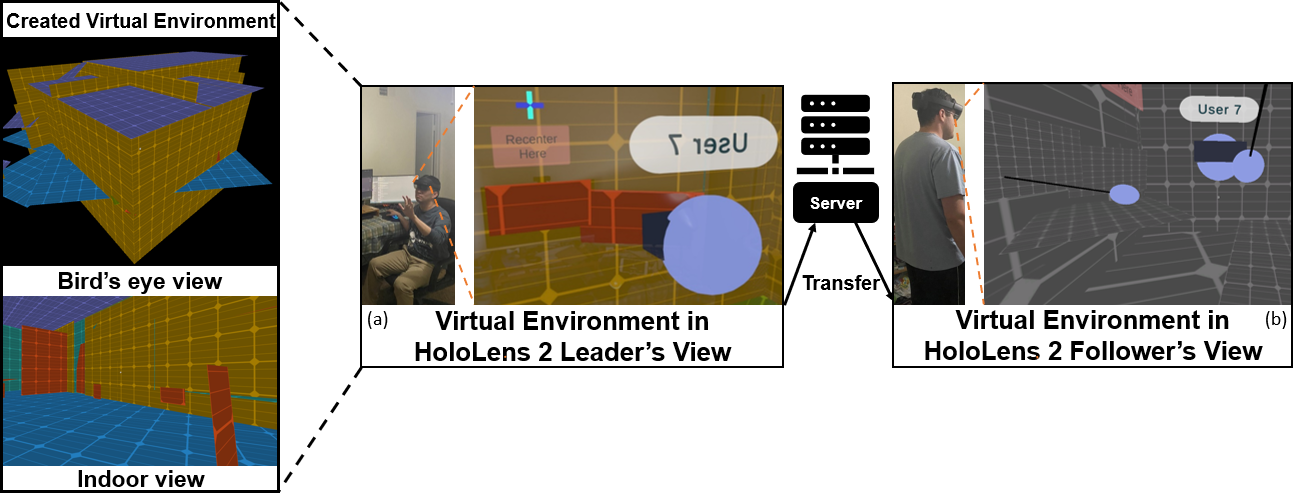}
\caption{Scene understanding generated virtual environment with network framework sharing between Leader and Followers with avatar indicating user's location. Different colors are used to indicate different categories of scene objects: Yellow indicates walls, red indicates medium-sized platforms, navy blue indicates ceilings, bright blue indicates floors, grass green indicates large-sized horizontal platforms, and blue-green indicates unclassified scene objects. (a) Bird's eye view and indoor view of virtual environment adapting to the physical environment created in Leader's device. (b) Follower's views of the MR environment with gray color scene objects indicate that the current user is not the creator.}
\label{fig:VirtualReal}
\vspace{-1em}
\end{figure*}

Scene understanding is a pre-built feature in the MRTK (Mixed Reality Tool-Kit from Microsoft) \cite{hololensMRTK} that is often used to observe and understand the target physical environment to obtain information and analyze it. This feature utilizes the spatial mapping feature of HoloLens 2, which uses a long-throw depth camera to capture the structure of the physical environment when users walk around and scan the surroundings to create multiple virtual flat planes (the virtual flat planes will be referred to as "scene objects" in the following articles) that align with the corresponding physical flat planes to create a complete MR virtual environment, as shown in Figure \ref{fig:VirtualReal} (a). 
In the CAMRE framework, we integrate this scene understanding feature to generate virtual environments that dynamically adapt to the changes in the Leader’s physical environment in close-to-real-time. 
Users can operate the virtual environment update settings on the control panel to manually update or auto-update with specific time intervals (5 seconds in default, can be changed).
The scene understanding feature in MRTK generates simple virtual planes to construct the virtual environment while preserving proper scene information.
As a result, the data size of the created virtual environment is relatively small (about 0.3 Megabytes) compared to standard virtual room mesh data (around 100 Megabytes). For instance, in a recent study \cite{Laskos2020}, an MR collaboration system was developed where user avatars were built and shared as mesh data, with the smallest data taking up 0.4 Megabytes, which is higher than the size of our scene understanding-based virtual environment.
Therefore, system load and environment creation time can decrease when a user moves and updates the virtual environment with low latency. 
In addition, these dynamically generated MR environments are shared with a set of Followers to facilitate collaboration. 
Whenever the Leader moves to or faces a new and unobserved physical environment, our system will update the virtual environment dynamically. Each scene object created or updated in the virtual environment on the Leader's HoloLens 2 will immediately update to the server and forward to all Followers when Leader creates it. Due to the small data size, the sharing process from Leader to Followers will have close-to-real-time latency. 
Followers can see the exact same MR environments in their devices to gain the same environmental information as the Leader; scene objects received from the Leader's side are shown as gray color only, which indicates that the current user is not the creator of these objects to avoid confusion, as shown in Figure \ref{fig:VirtualReal} (b).

\subsubsection{Network Framework}\hspace{\fill}\\
In CAMRE, we used two state-of-the-art Unity networking frameworks, including Unity Netcode for Gameobject with Unity Relay and Photon Unity Networking, to transfer observed scene objects from Leader to an external server and then to Followers to accomplish remote collaboration. 
We have built the CAMRE system on two different frameworks to compare the network latency and ease of use, which will be evaluated in a later section. 
\begin{enumerate}
\vspace{-2mm}
\item \textbf{Netcode for Gameobject} is the latest (first released in June 2022) and highly recommended package built by Unity for multiplayer networking that enables the system to synchronize virtual objects' position, rotation, and scale. 
Since Netcode for GameObject only supports local network connections without requiring modification of the user's router, to avoid complicated operations to ensure user-friendliness, we use the Unity Relay, a Unity-registered third-party package, for external network connections. The combined system allows up to 50 concurrent users for free (\$0.16 per additional user) but requires Leader to send access codes to Followers externally.
\vspace{-5mm}
\item Another networking framework we used is \textbf{Photon Unity Networking}, a primarily recommended multiplayer network framework for HoloLens 2 to perform multi-user collaboration. This system offers similar functionality as it allows virtual objects' position, rotation, and scale sharing through Photon server, which allows users to join a preset room without exchanging external messages. However, the free version of the system supports only 20 concurrent users. (up to 2000 concurrent users for \$370).
\vspace{-2mm}
\end{enumerate}

\subsection{CAMRE's Expanded Navigation Features}\label{CAMRE_Expanded}
In this section, we describe three expanded features of CAMRE to provide an overview and depth perception of the virtual environment that can assist in user navigation and understanding of the entire environment. 
Here, we make the following two assumptions for users before starting to use the expanded navigation features:
\begin{enumerate}
    \vspace{-2mm}
    \item CAMRE MR virtual environment is observed and built completely by the HoloLens 2.
    \vspace{-2mm}
    \item Information behind the obstacles is available to the user.
    \vspace{-2mm}
\end{enumerate}

\subsubsection{Dynamic X-ray Vision Window}\hspace{\fill}\\
In order to provide additional information and depth perception (such as motion parallax) of the surrounding scene to the users, we built a dynamic X-ray vision window \cite{guo2022dynamic}. This feature allows users to directly see through the obstacles in front of them in the CAMRE MR virtual environment while still retaining a complete view of the surrounding environment to obtain information behind obstacles by utilizing the clipping primitive feature in MRTK. By attaching the clipping primitive onto selected virtual objects to mimic a physical window and make the contact area partially transparent, users have the ability to gain information behind obstacles before physically moving to other rooms. 
To provide customization and avoid potential motion sickness, users can dynamically change the X-ray vision window's size with a slider to best fit their current viewing needs.
Furthermore, we used the eye-tracking function in HoloLens 2 to allow the X-ray vision window to follow the eye-gaze direction, dynamically updating the window and making it move smoothly and quickly. Having the X-ray vision window updates based on the user's position, movement, and eye gaze can provide motion parallax to more closely resemble the real world. 
To prevent users from experiencing virtual motion sickness while using the eye gaze X-ray vision window, we offer an alternative option, head gaze version (window following head movement), that follows head movement, which enables users to choose the version they are comfortable with.
With the help of X-ray vision in the CAMRE MR virtual environment, users can locate and perceive the distance of objects in adjacent rooms without physically moving. A depiction of this feature is shown in Figure \ref{fig:XRayAndWallTrans} (a).



\subsubsection{Complete See-through Virtual Environment}
\begin{figure*}[htbp]
\centering
\includegraphics[width=\textwidth, keepaspectratio=true]{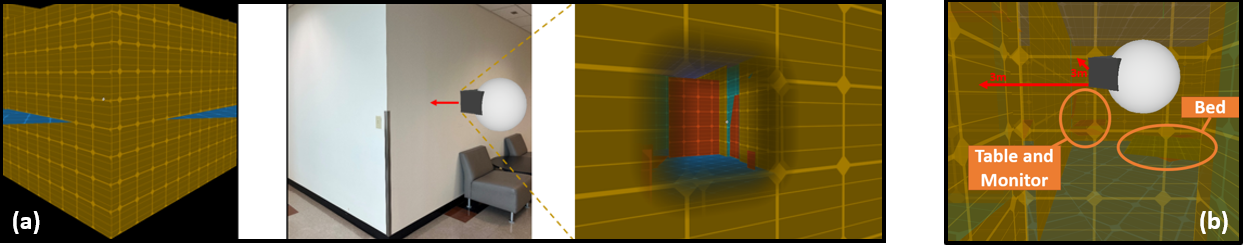}
\caption{(a) Example dynamic X-ray vision window view on virtual environment (b) Complete see-through virtual environment to make virtual wall objects transparent to help users see objects inside the room when users move approach within 3 meters.}
\label{fig:XRayAndWallTrans}
\vspace{-2em}
\end{figure*}

\hspace{\fill}\\
We also provided an option for the users to have a complete direct view of the created MR virtual environment when they navigate the surroundings. 
As users approach the virtual wall objects created by CAMRE's scene understanding feature within a 3-meter radius (euclidean distance between the user's current location and wall object's location), the objects become 30\% transparent, which allows users to view information about adjacent rooms before leaving the current one and also helps them perceive the distance between themselves and the edges of the virtual environment. Conversely, when the user moves away from the virtual wall objects beyond three meters, the objects will return to opaque. 
We ensure that users are aware of significant changes in the virtual environment by alerting them using spatial sound cues from the direction of wall objects they approach and becoming transparent. For instance, if a user moves towards a wall object on the right side, an alert sound will be heard on the right side to indicate the approaching movement.
By generating sound cues from changing objects using the HoloLens 2's spatial sound capabilities, users can quickly notice where wall objects are within three meters and have changed. This can help the users acquire spatial information for additional depth cues while obtaining the information behind the obstacles. The wall object's transparent effect is shown in Figure \ref{fig:XRayAndWallTrans} (b).


\subsubsection{Real-time Mini-map}
\begin{figure}[htbp]
\centering
\includegraphics[width=0.45\textwidth, keepaspectratio=true]{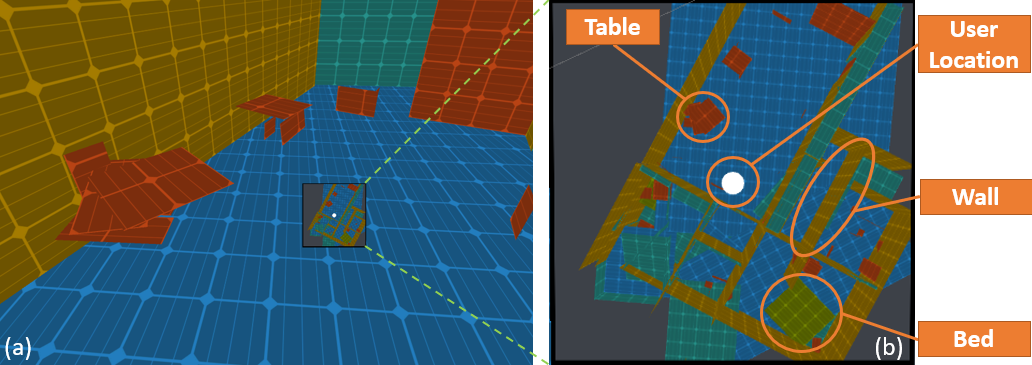}
\caption{(a) Mini-map position in user's FOV and (b) larger display to show user's location and scene objects captured by CAMRE on mini-map.}
\label{fig:MiniAvatar}
\vspace{-1em}
\end{figure}

\hspace{\fill}\\
We also include a mini-map capability to provide a bird's eye view of the entire CAMRE MR virtual environment to gain an overall understanding. Our assumption here is that the Leader's CAMRE MR virtual environment is observed and built completely beforehand and shared with the Followers. This allows Followers to explore the entire virtual environment independently with the real-time mini-map without following the Leader in real-time. This type of mini-map is a common feature in first-person shooter video games to help players navigate their surroundings. Similarly, including a mini-map feature in the CAMRE framework can allow users to maintain an awareness and understanding of their surrounding environment. 
Therefore, we create a track-up (this setting is a configurable option, where a north-up mini-map can also be chosen) mini-map that updates in real-time with the user's physical movement (position and rotation) and will follow and display at the bottom right corner of the user's FOV, as shown in Figure \ref{fig:MiniAvatar} (a).
To identify users on the mini-map, we create a self-avatar following the user’s position in real-time, shown on the mini-map to indicate the current position. Avatar on the Leader side will spawn at the origin point where the Leader starts the application, and avatars on Followers side will also spawn at the Leader's origin point whenever connected to the server. All users can locate the current location of other users to confirm whether the virtual object being shared is within the other user's FOV.
Examples of the mini-map avatar and virtual objects displayed on the mini-map are shown in Figure \ref{fig:MiniAvatar} (b).

\begin{figure}[htbp]
\centering
\includegraphics[width=0.45\textwidth, keepaspectratio=true]{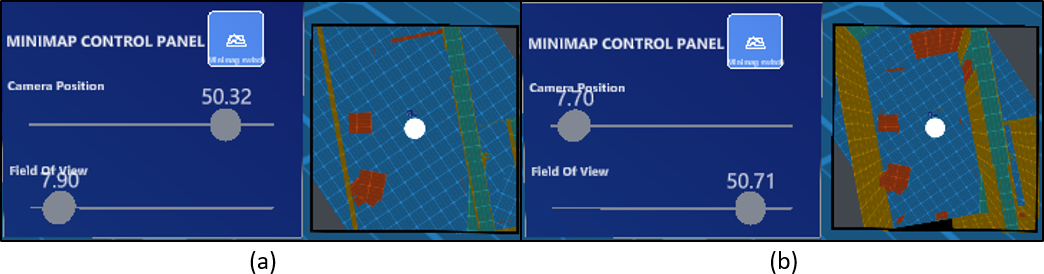}
\caption{(a) high camera position and low FOV result in a flatter floor plan, (b) low camera position and high FOV result in higher perspective}
\label{fig:MiniControl}
\vspace{-1em}
\end{figure}

Our system offers users the ability to control the position of the camera and the field of view of the mini-map in real-time. Through two sliders, users can choose to view a close-up of a specific area to display detailed information or a full view of the entire CAMRE MR virtual environment to gather complete information about other rooms before physically moving to them. 
Furthermore, by combining different settings of the two sliders, the mini-map can display varying levels of detail to provide further information.
Suppose the user chooses a high camera position value and a low FOV value. In this case, the mini-map will display a flatter and complete floor plan to help users see the top view of the virtual objects located in the surrounding virtual environment and avoid scene objects (such as wall objects) affecting the judgment of the virtual object's position, as shown in Figure \ref{fig:MiniControl} (a).
Conversely, if the user chooses a low camera position value and a high FOV value, the mini-map will present a higher perspective of the scene objects, allowing the user to see the three-dimensional view of the scene objects more clearly to help users locate and calculate the size of the scene objects, as shown in Figure \ref{fig:MiniControl} (b).

By combining the low-latency environmental update attribute of our CAMRE system, the above three expanded features can provide additional information for the users (including adjacent room settings and depth cues) when they physically move in the created virtual environment based on the surrounding physical environment. Typically, users have a compressed depth perception when immersed in a virtual environment; this may be partially due to the lack of certain types of contextual information typical of the physical environment (such as shadows of physical objects). 
By including the above three expanded features, CAMRE can help users to move smoothly to the desired location apart from providing depth perception to enhance immersion.

\section{CAMRE System Evaluation Experiment Design}\label{section-experimental-design}
In this section, we designed multiple experiments to evaluate the latency of each feature to provide the capability details of the CAMRE system, including time taken to construct a virtual environment, scene object transfer latency and throughput (average bytes transferred per second), transfer packet loss, latency for X-ray vision, and latency for mini-map.
The major factors we used for experiment design are:
\begin{itemize}
    \vspace{-2mm}
    \item Room size difference.
    \vspace{-2mm}
    \item Testing different networking frameworks.
    \vspace{-2mm}
    \item Number of contemporary connections.
    \vspace{-2mm}
    \item Distance between Leader and Followers.
    \vspace{-2mm}
\end{itemize}

\subsection{CAMRE MR Virtual Environment Evaluation}

\subsubsection{Virtual Environment Data Scale and Constructing Time}\hspace{\fill}\\
To comprehend the magnitude of data and time required for users to create and explore the virtual environment, we assessed the data size and construction time of Leader's CAMRE MR virtual environment. We conducted this evaluation by incorporating three rooms of varying sizes: a personal room (3.81m X 3.02m X 2.40m, containing approximately 30 scene objects), a living room (7m X 3.92m X 2.97m, containing approximately 90 scene objects), and a large classroom (13m x 9.2m x 3m, containing approximately 130 scene objects). This evaluation aimed to determine if the size of the room has an impact on the data size and construction time.
During this experiment, we record the time span from pressing the "update scene" button to spawning the last scene object by directly recording the system timer, and we do the experiment 20 times for each room to account for any variation that might occur. 


\subsubsection{Virtual Environment Transfer Latency}\hspace{\fill}\\
In CAMRE, we used Unity Netcode for Gameobject with Unity Relay or Photon Unity Networking to transfer the observed virtual environment between multiple devices. Therefore, to compare the pros and cons of the two selected networking frameworks, we measure them by using the following metrics:
\begin{itemize}
    \vspace{-2mm}
    \item \textbf{Leader to Follower Data transfer latency/Standard Deviation}: average time difference between the Leader side creating each scene object and the Follower side receiving the scene object with the standard deviation across all observed time differences. 
    \vspace{-2mm}
    \item \textbf{Room size scene (50 scene objects) transfer time}: During the experiment, we capture the average transfer time as the time difference between receiving the first and last scene objects with average bytes transferred from Leader to Follower and the total number of transferred scene objects as a benchmark. To ensure a fair comparison of transfer times among different Leaders while accounting for varying room sizes, we use the following equation to normalize the transfer time to a standard virtual room with 50-scene objects (a standard indoor room size according to other Leaders' created virtual environment in our medium distance scenario): 
    \vspace{-2mm}
\[Total Transfer Time / Total Scene Objects * 50\]
    \vspace{-8mm}
    \item \textbf{Average throughput(bytes per second)}: average bytes received per second on the Follower side during the process of transferring the entire virtual environment.
    \vspace{-2mm}
    \item \textbf{Packet loss}: packet loss percentage over the whole virtual environment transfer. 
    \vspace{-2mm}
\end{itemize}
In this experiment, we investigated whether the number of concurrently connected users on the same server and the distance between the Leader and Followers affect transfer efficiency. Therefore, we divide the experiment into three different scenarios (each experiment is conducted 5 times to account for any variation); detailed settings are listed in Table \ref{table:scenario}:


\begin{table}[htbp]
\caption{Experiment Scenarios}
\vspace{-1em}
\begin{center}
\includegraphics[width=0.45\textwidth, keepaspectratio=true]{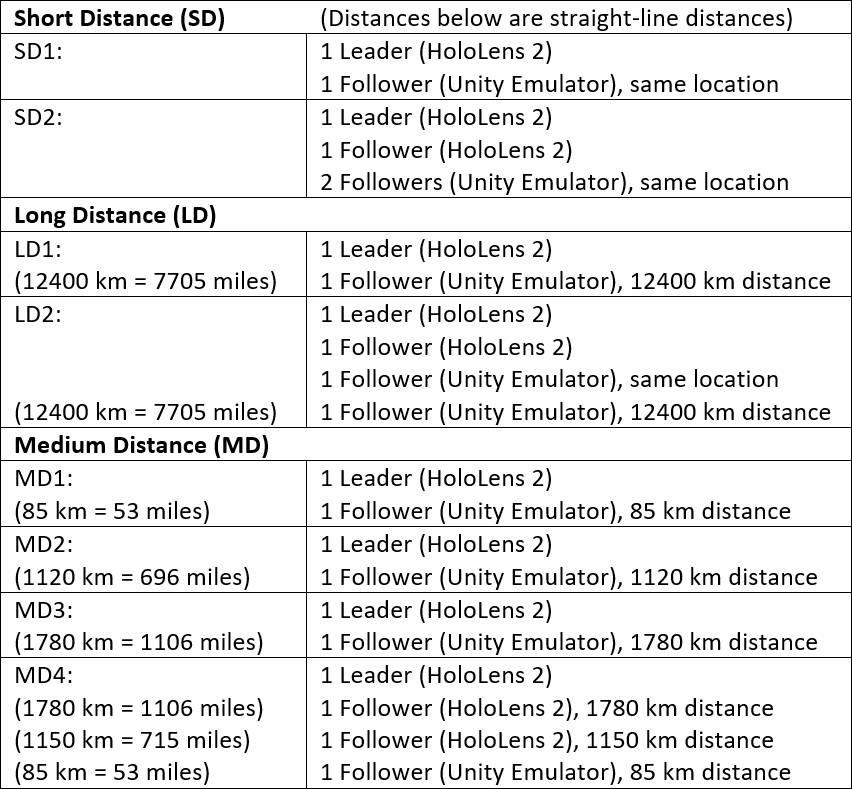}
\label{table:scenario}
\vspace{-2em}
\end{center}
\end{table}

To capture packet data transferring from Leader to Follower, we set up one Follower using a Unity emulator as the main evaluation target and used Wireshark \cite{WiresharkWeb} to capture the data. 
We do each experiment combination 5 times to share scene objects to account for any variation that might occur.


\subsection{CAMRE Expanded Features Evaluation}
In addition to evaluating the base CAMRE MR virtual environment, we also assess the expanded features for real-time expression.
The complete see-through virtual environment function is primarily designed to adjust the transparency of virtual wall objects when the user approaches the wall within 3 meters. However, due to the short latency of the transparency adjustment and the natural slight movement of the user's head while walking and wearing the HoloLens 2, it becomes challenging to precisely measure the distance accuracy in millimeter-scale between the user and the wall object from an external perspective. 
Therefore, in this subsection, we will only evaluate dynamic X-ray vision and mini-map features.

\subsubsection{Dynamic X-ray Vision Window Display Latency}\hspace{\fill}\\
We conducted an evaluation to determine if the dynamic X-ray vision window has low latency, providing users with a real-time experience. To calculate the display latency, we recorded the timestamp of when the X-ray vision enabling switch was pressed and when the X-ray vision window was displayed, measuring the time gap between them.
This experiment is repeated 100 times to account for any variation that might occur.

\subsubsection{Dynamic X-ray Vision Window Moving Latency}\hspace{\fill}\\
To evaluate whether the dynamic X-ray vision window consistently follows the user's eye-gaze direction when physically moving in the CAMRE MR virtual environment, we evaluate the moving latency of the dynamic X-ray vision window by calculating the time difference between the user's eye-gaze movement captured by HoloLens 2 (direction can be digitized using eye-gaze ray intersections with wall objects) and the time X-ray vision window position updated to the same position by recording the timestamp.
This experiment is also repeated 100 times to account for any variation that might occur.

\subsubsection{Mini-map Moving Latency}\hspace{\fill}\\
Consistently following the user's movement is essential for the Mini-map to help determine their current location in the virtual environment and related locations from other collaborators since Followers can move independently without following the Leader in real-time. Therefore, we conducted an evaluation to determine if the mini-map accurately tracks the user's physical movements.
Specifically, we measured the mini-map moving latency by recording the timestamp to calculate the time difference between when the user rotated the display by 180 degrees (ensuring that there is a noticeable angle difference between the direction displayed by the mini-map and the user's current facing direction) and when the mini-map rotation caught up to the same rotation.
This experiment is also repeated 100 times to account for any variation that might occur.

\section{CAMRE Evaluation Results and Discussion}\label{section-results}
With the evaluation experiments proposed in the previous section, we collected various experimental results to further analyze and discuss in detail to the performance of CAMRE. 

\subsection{CAMRE MR Virtual Environment Data Size and Construction Time}
In this experiment, we built three different-sized rooms into 30, 90, and 130 scene objects to explore virtual environment construction time. 
By saving three different scene objects into a bytes file, their respective data sizes are in order: 0.18, 0.33, 0.75 megabytes.
According to the experiment results in Table \ref{table:VEBuildingTime}, the personal room took an average of 0.96 seconds with a standard deviation of 0.13 to fully build the virtual environment, the living room took an average of 2.53 seconds with a standard deviation of 0.28, and large classroom took average 3.69 seconds with a standard deviation of 0.10, which shows a low construction time attribute to build a complete 3D indoor environment comparing to \cite{Li2020} with 12 seconds and \cite{zhura2023neuroswarm} with a 60 seconds indoor scene to 3D mesh reconstruction computation time. 
Low data size and construction time can further benefit CAMRE updating and streaming virtual information between Leader and Followers when Leader physically moves in the surrounding environment.

\begin{table}[h]
\caption{CAMRE MR Virtual Environment Construction Time}
 \begin{tabular}{| c | c | c | c | c | c |} 
 \hline
 & Personal & Living & Large & \cite{Li2020} & \cite{zhura2023neuroswarm}\\ 
 & Room & Room & Classroom & & \\
 \hline
 Time (s) & 0.96s & 2.53s & 3.69s & 12s & 60s\\
 \hline
\end{tabular}
\label{table:VEBuildingTime} 
\vspace{-2em}
\end{table}


\subsection{CAMRE MR Virtual Environment Transfer Latency}
To evaluate and prove the low transfer latency sharing virtual environments with two different networking frameworks, we conducted three experiment scenarios with different distances and different numbers of concurrent users between Leaders and Followers. 
We also consider internet bandwidth as a potential factor affecting transfer latency; therefore, we asked all Leaders and Followers to provide their internet bandwidth showing in \url{https://fast.com/}.
Before starting the experiments, we calibrated all connected machines with Network Time Protocol (NTP) to confirm the accuracy of the latency with milliseconds preciseness. 
Before the primary observing Follower connects to the server, we launched Wireshark to capture internet packets and stop capturing after receiving all the scene objects transferred from the Leader. 
Since HoloLens 2 does not provide software for users to capture internet packet data, we can only capture and analyze packet data transferred from networking servers in the Unity emulator Follower side.  
According to the packet data, both Photon networking and Netcode for Gameobject use User Datagram Protocol (UDP) to transfer data, and there is no packet loss in every scenario discussed in the following. 

\subsubsection{Scenario 1: Short Distance (SD)}\hspace{\fill}\\
The Leader and Followers in this scenario all physically stay in the same location with an internet speed of 240 Mb per second. 
The data shown in Table \ref{table:SDLD} indicates that there is no significant difference in the three evaluation metrics when connecting with one Leader and one Follower (SD1) and one Leader and three Followers (SD2) under the two networking frameworks, which reflects the low-latency stability of the system even when connecting with multiple users.
During our testing, we discovered that when transferring a room-sized scene, Netcode for Gameobject took longer than Photon networking, but had a higher throughput. Our analysis of the captured packet data revealed that Netcode encrypts the transferred data, which could lead to larger data size, while Photon transfers data directly.

\begin{table}[htbp]
\caption{Short Distance (SD) and Long Distance (LD) Scenario}
\vspace{-1em}
\begin{center}
\includegraphics[width=0.4\textwidth, keepaspectratio=true]{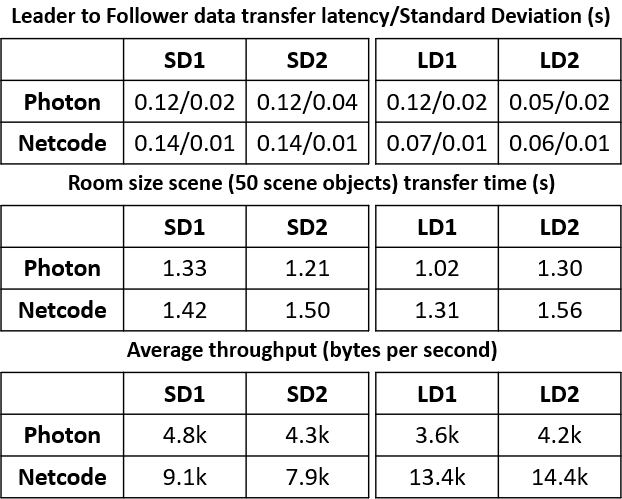}
\label{table:SDLD}
\vspace{-2em}
\end{center}
\end{table}

\subsubsection{Scenario 2: Long Distance (LD)}\hspace{\fill}\\
In this scenario, only the primary observing Follower is located at another place with about 12,400 km distance with 530 Mbps internet speed, Leader and the other two Followers are located at the same location with 240 Mbps internet speed. 
Based on Table \ref{table:SDLD}, connecting a Leader with one Follower (LD1) and three Followers (LD2) shows no significant difference in the data transfer latency category, still exhibiting a stable attribute for each scene object.
However, transferring a room-sized scene with the LD2 scenario takes a little longer to complete the process, which means that if there is only one user located farther away from the location where other users are located, the scene transfer time will be affected.
Furthermore, the average throughput in LD1 and LD2 scenarios while using Photon networking displays no significant difference when compared to the short-distance scenario. However, Netcode for Gameobject exhibits a higher throughput, suggesting that internet speed might affect the throughput of Netcode, but does not provide a significant difference for the Photon networking framework.

\begin{table}[htbp]
\caption{Medium Distance (MD) Scenario}
\vspace{-2em}
\begin{center}
\includegraphics[width=0.48\textwidth, keepaspectratio=true]{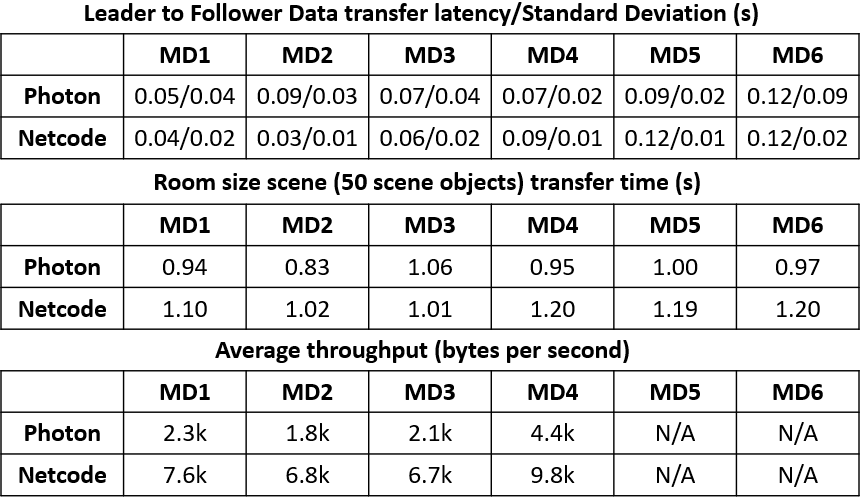}
\label{table:Medium}
\vspace{-3em}
\end{center}
\end{table}

\subsubsection{Scenario 3: Medium Distance (MD)}\hspace{\fill}\\
In this scenario, Leader and all other Followers are located at different locations. MD1, MD2, and MD3 have different Leaders with internet speeds in order: 90, 250, and 90 Mbps, and the Follower is the same with 240 Mbps internet speed. MD4, MD5, and MD6 have the Leader with 90 Mbps internet speed and other Followers with 240, 250, and 90 Mbps in order.
The experiment results shown in Table \ref{table:Medium} indicate that data transfer latency is slightly higher in MD4, MD5, and MD6 scenarios compared to MD1, MD2, and MD3 scenarios respectively. Similarly, we observe a similar pattern when transferring a room-sized scene over the Netcode server, suggesting that connecting from multiple locations may impact the transfer performance.
Due to the use of HoloLens 2 devices by the other two Followers to connect with Leader, we were unable to capture internet packets for MD5 and MD6 scenarios; therefore, we marked N/A to MD5 and MD6 in the average throughput section of Table \ref{table:Medium}. 
After analyzing the average throughput results, we observed that MD4 has a higher throughput compared to MD1, MD2, and MD3 scenarios, which could indicate that Photon and Netcode servers require higher throughput to efficiently transfer data with multiple users across different locations.

Based on the above experiment results, all of the Leader to Follower data transfer latencies are sub-0.15 seconds, and room size scene transfer time is lower than 1.6 seconds in both networking frameworks, which indicates that the CAMRE system can transfer 3D virtual environments with low latency by utilizing small data size to increase collaboration consistency between Leader and Followers. 

\begin{table}[htbp]
\caption{Networking Framework Comparison Baseline}
\vspace{-2em}
\begin{center}
\includegraphics[width=0.48\textwidth, keepaspectratio=true]{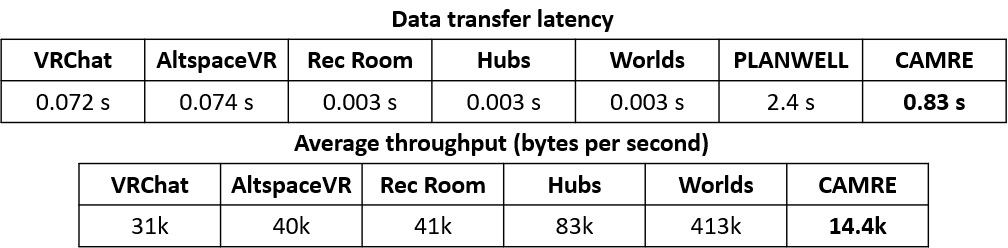}
\label{table:Comparison}
\vspace{-3em}
\end{center}
\end{table}

\subsubsection{Comparison with Existing Systems}
We also provide comparisons with other existing collaboration systems; however, only one AR collaboration system, PLANWELL \cite{Nittala2015}, measures data transfer latency. Therefore, we found multiple state-of-the-art VR multiuser platforms, including VRChat \cite{VRChat}, AltspaceVR \cite{AltspaceVR}, Rec Room \cite{RecRoom}, Mozilla Hubs \cite{MozillaHubs}, and Horizon Worlds \cite{HorizonWorlds}, that transfer data to multiple users, which are measured in \cite{Cheng2022} as client-to-server and back-to-client round-trip time that is similar to a Leader transferring packets to a server and then to a Follower. 
The comparison table (Table \ref{table:Comparison}) shows that PLANWELL has a data transfer latency of 2.4 seconds, which is higher than the latency of CAMRE in any scenario. All other VR systems have better data transfer latency than the CAMRE system. 
However, it is important to note that CAMRE is an MR collaboration system that requires capturing information from the physical environment, which could increase system load. The fact that CAMRE has similar data transfer latency as two other VR systems indicates that its low data size design has a significant impact.
In addition, \cite{Cheng2022} also measured the average throughput of the five state-of-the-art VR collaboration systems; therefore, we have compared the average throughput of our CAMRE system with other VR collaboration systems. Even though the two network frameworks we have employed have lesser throughput than the state-of-the-art VR systems, the CAMRE system archives low data transfer latency, as the server resources we possess are not as high as those used by large companies.

\subsection{Dynamic X-ray Vision Window Display Latency}
\begin{figure}[htbp]
\centering
\includegraphics[width=0.4\textwidth, keepaspectratio=true]{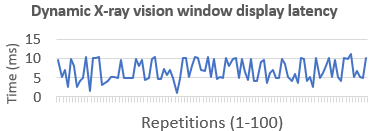}
\caption{Dynamic X-ray vision window display latency in milliseconds that is repeated 100 times to account for any variation that might occur. (Average: 6.81 milliseconds with standard deviation: 2.63)}
\label{fig:XRayDisplayLatency}
\vspace{-1em}
\end{figure}

During the experiment process, we conducted the switch-enabling procedure 100 times. The results are displayed in Figure \ref{fig:XRayDisplayLatency}. The average display time was 6.81 milliseconds, with a standard deviation of 2.63 milliseconds. 
Our findings indicate that the latency period for the X-ray vision window to appear on the display after the switch is pressed is consistently small enough to be considered a real-time feature, which improves usability and reduces the likelihood of virtual motion sickness \cite{stauffert2020latency}.

\subsection{Dynamic X-ray Vision Window Moving Latency}
\begin{figure}[htbp]
\centering
\includegraphics[width=0.4\textwidth, keepaspectratio=true]{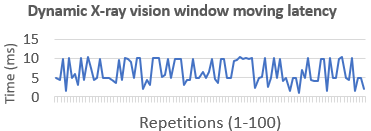}
\caption{Dynamic X-ray vision window moving latency in milliseconds that is repeated 100 times to account for any variation that might occur. (Average: 6.57 milliseconds with standard deviation: 2.92)}
\label{fig:XRayMovingLatency}
\vspace{-1em}
\end{figure}

Similarly, we conducted the eye-gaze movement (move 45 degrees from left to right) process 100 times, and the results are shown in Figure \ref{fig:XRayMovingLatency}. The average displaying time was 6.57 milliseconds, with a standard deviation of 2.92 milliseconds. 
Results show a consistently low latency that can be considered a real-time feature that instantly provides information inside adjacent rooms when users move their eye-gaze direction. 
By utilizing the physical window-like X-ray vision that follows the user's eye-gaze direction in real-time, the X-ray vision window is capable of providing motion parallax, a feature that is normally available in a real environment but is unavailable in the virtual environment.

\subsection{Mini-map Moving Latency}
\begin{figure}[htbp]
\centering
\includegraphics[width=0.4\textwidth, keepaspectratio=true]{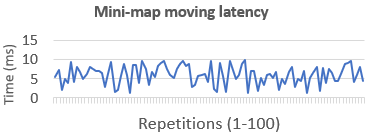}
\caption{Mini-map moving latency in milliseconds that is repeated 100 times to account for any variation that might occur. (Average: 5.99 milliseconds with standard deviation: 2.42)}
\label{fig:MiniMovingLatency}
\vspace{-1em}
\end{figure}

We evaluate this feature by performing the rotation task 100 times, and the results are shown in Figure \ref{fig:MiniMovingLatency}.
The average displaying latency was 5.99 milliseconds, with a standard deviation of 2.42 milliseconds. 
Results indicate that this feature has a consistent latency of under 10 milliseconds, which could show the real-time attribute of this feature. Having a mini-map displays the user's surrounding virtual environment in real-time, and with the camera options (Figure \ref{fig:MiniControl}) that allow the user to gain a broader view of the environment, users can quickly locate the location of the target virtual objects. 

\section{Conclusion and Future Work}\label{section-conclusion}
In the CAMRE framework, we demonstrate dynamically generated MR virtual environments with low latency and small data sizes in the HoloLens 2 device by utilizing the scene understanding feature. 
With the small data sizes of the virtual environment, we employ a Leader-Follower paradigm, representing the Leader's surrounding physical environment and transferring these environments through networking frameworks to the Followers in close-to-real-time.
This permits remote connections and collaboration, including real-time expanded features to assist users with navigation.  
As part of our research, we evaluated the performance of the CAMRE framework, which showed that users can construct virtual environments in a short amount of time. Our tests revealed that it takes around 2.5 seconds to build a living room using the framework. We also evaluated two networking frameworks for sharing a typical room size scene and found that their latencies were below 1.6 seconds in all evaluated scenarios. We then assessed the X-ray vision and mini-map display and found that their update latencies were all below 12 milliseconds, which suggests that these features can be used in real-time to help users navigate through the virtual environment.

In the future, to address the limitations described in Section \ref{section-limitations}, we plan to design and conduct an exhaustive set of behavioral studies to understand how users perceive CAMRE as a means of collaboration as well as the efficacy of the MR navigation features. Furthermore, we would also like to investigate the networking performance of CAMRE with more than four concurrent users.
Currently, we do not allow the roles of Leader and Followers to be swapped in real-time, but we plan to investigate such a role swapping in the future.
Lastly, we aim to enhance the collaborative capabilities of the CAMRE system by implementing real-time frame analysis for the creation of virtual object mesh and dynamic color adaptation of virtual objects to their surrounding environment. At present, the system generates a virtual environment with basic geometric scene objects. However, by fitting primitive geometry \cite{Li2011GlobFit, alghofaili2023warpy} to these objects, we could potentially create more detailed virtual objects without overburdening computational resources.
These new features may further increase the effectiveness of the CAMRE framework.

\acknowledgments{
This research was sponsored by the DEVCOM U.S. Army Research Laboratory under Cooperative Agreement Number W911NF-21-2-0145 to B.P. 
\\
The views and conclusions contained in this document are those of the authors and should not be interpreted as representing the official policies, either expressed or implied, of the DEVCOM Army Research Laboratory or the U.S. Government. The U.S. Government is authorized to reproduce and distribute reprints for Government purposes notwithstanding any copyright notation.}


\bibliographystyle{abbrv-doi}

\bibliography{CAARE_IEEEVR}

\begin{thebibliography}{10}

\bibitem{alghofaili2023warpy}
R.~Alghofaili, C.~Nguyen, V.~Krs, N.~Carr, R.~M{\u{e}}ch, and L.-F. Yu.
\newblock Warpy: Sketching environment-aware 3d curves in mobile augmented reality.
\newblock In {\em 2023 IEEE Conference Virtual Reality and 3D User Interfaces (VR)}, pp. 367--377. IEEE, 2023.

\bibitem{guo2022dynamic}
A.~Anonymous.
\newblock Hidden title.
\newblock In {\em Hidden}, p. Hidden, Hidden. doi: {{%
Hidden}}


\bibitem{Benford1998}
S.~Benford, C.~Greenhalgh, G.~Reynard, C.~Brown, and B.~Koleva.
\newblock Understanding and constructing shared spaces with mixed-reality boundaries.
\newblock {\em ACM Trans. Comput.-Hum. Interact.}, 5(3):185–223, sep 1998. doi: {{%
10\hspace{.1pt}\discretionary{.}{%
}{.}\hspace{.4pt}1145\discretionary{/}{%
}{/}292834\hspace{.1pt}\discretionary{.}{%
}{.}\hspace{.4pt}292836}}


\bibitem{Cheng2022}
R.~Cheng, N.~Wu, M.~Varvello, S.~Chen, and B.~Han.
\newblock Are we ready for metaverse? a measurement study of social virtual reality platforms.
\newblock In {\em Proceedings of the 22nd ACM Internet Measurement Conference}, IMC '22, p. 504–518. Association for Computing Machinery, New York, NY, USA, 2022. doi: {{%
10\hspace{.1pt}\discretionary{.}{%
}{.}\hspace{.4pt}1145\discretionary{/}{%
}{/}3517745\hspace{.1pt}\discretionary{.}{%
}{.}\hspace{.4pt}3561417}}


\bibitem{Druta2021}
R.~Druta, C.~Druta, P.~Negirla, and I.~Silea.
\newblock A review on methods and systems for remote collaboration.
\newblock {\em Applied Sciences}, 11(21), 2021. doi: {{%
10\hspace{.1pt}\discretionary{.}{%
}{.}\hspace{.4pt}3390\discretionary{/}{%
}{/}app112110035}}


\bibitem{Duval2014}
T.~Duval, T.~T.~H. Nguyen, C.~Fleury, A.~Chauffaut, G.~Dumont, and V.~Gouranton.
\newblock Improving awareness for 3d virtual collaboration by embedding the features of users’ physical environments and by augmenting interaction tools with cognitive feedback cues.
\newblock {\em Journal on Multimodal User Interfaces}, 8(2):187--197, 2014.

\bibitem{Fan2022}
J.~Fan, P.~Zheng, and S.~Li.
\newblock Vision-based holistic scene understanding towards proactive human–robot collaboration.
\newblock {\em Robotics and Computer-Integrated Manufacturing}, 75:102304, 2022. doi: {{%
10\hspace{.1pt}\discretionary{.}{%
}{.}\hspace{.4pt}1016\discretionary{/}{%
}{/}j\hspace{.1pt}\discretionary{.}{%
}{.}\hspace{.4pt}rcim\hspace{.1pt}\discretionary{.}{%
}{.}\hspace{.4pt}2021\hspace{.1pt}\discretionary{.}{%
}{.}\hspace{.4pt}102304}}


\bibitem{Jones2008}
A.~Jones, J.~E. Swan, G.~Singh, and E.~Kolstad.
\newblock The effects of virtual reality, augmented reality, and motion parallax on egocentric depth perception.
\newblock In {\em 2008 IEEE Virtual Reality Conference}, pp. 267--268, 2008. doi: {{%
10\hspace{.1pt}\discretionary{.}{%
}{.}\hspace{.4pt}1109\discretionary{/}{%
}{/}VR\hspace{.1pt}\discretionary{.}{%
}{.}\hspace{.4pt}2008\hspace{.1pt}\discretionary{.}{%
}{.}\hspace{.4pt}4480794}}


\bibitem{Jones2011}
J.~A. Jones, J.~E. Swan, G.~Singh, and S.~R. Ellis.
\newblock Peripheral visual information and its effect on distance judgments in virtual and augmented environments.
\newblock In {\em Proceedings of the ACM SIGGRAPH Symposium on Applied Perception in Graphics and Visualization}, APGV '11, p. 29–36. Association for Computing Machinery, New York, NY, USA, 2011. doi: {{%
10\hspace{.1pt}\discretionary{.}{%
}{.}\hspace{.4pt}1145\discretionary{/}{%
}{/}2077451\hspace{.1pt}\discretionary{.}{%
}{.}\hspace{.4pt}2077457}}


\bibitem{Laskos2020}
D.~Laskos and K.~Moustakas.
\newblock Real-time upper body reconstruction and streaming for mixed reality applications.
\newblock In {\em 2020 International Conference on Cyberworlds (CW)}, pp. 129--132, 2020. doi: {{%
10\hspace{.1pt}\discretionary{.}{%
}{.}\hspace{.4pt}1109\discretionary{/}{%
}{/}CW49994\hspace{.1pt}\discretionary{.}{%
}{.}\hspace{.4pt}2020\hspace{.1pt}\discretionary{.}{%
}{.}\hspace{.4pt}00027}}


\bibitem{Li2020}
C.~Li, L.~Yu, and S.~Fei.
\newblock Large-scale, real-time 3d scene reconstruction using visual and imu sensors.
\newblock {\em IEEE Sensors Journal}, 20(10):5597--5605, 2020. doi: {{%
10\hspace{.1pt}\discretionary{.}{%
}{.}\hspace{.4pt}1109\discretionary{/}{%
}{/}JSEN\hspace{.1pt}\discretionary{.}{%
}{.}\hspace{.4pt}2020\hspace{.1pt}\discretionary{.}{%
}{.}\hspace{.4pt}2971521}}


\bibitem{Li2011GlobFit}
Y.~Li, X.~Wu, Y.~Chrysathou, A.~Sharf, D.~Cohen-Or, and N.~J. Mitra.
\newblock Globfit: Consistently fitting primitives by discovering global relations.
\newblock 30(4), jul 2011. doi: {{%
10\hspace{.1pt}\discretionary{.}{%
}{.}\hspace{.4pt}1145\discretionary{/}{%
}{/}2010324\hspace{.1pt}\discretionary{.}{%
}{.}\hspace{.4pt}1964947}}


\bibitem{lugrin2016breaking}
J.-L. Lugrin, M.~E. Latoschik, M.~Habel, D.~Roth, C.~Seufert, and S.~Grafe.
\newblock Breaking bad behaviors: A new tool for learning classroom management using virtual reality.
\newblock {\em Frontiers in ICT}, 3:26, 2016.

\bibitem{Luo2007}
X.~Luo, R.~Kenyon, D.~Kamper, D.~Sandin, and T.~DeFanti.
\newblock The effects of scene complexity, stereovision, and motion parallax on size constancy in a virtual environment.
\newblock In {\em 2007 IEEE Virtual Reality Conference}, pp. 59--66, 2007. doi: {{%
10\hspace{.1pt}\discretionary{.}{%
}{.}\hspace{.4pt}1109\discretionary{/}{%
}{/}VR\hspace{.1pt}\discretionary{.}{%
}{.}\hspace{.4pt}2007\hspace{.1pt}\discretionary{.}{%
}{.}\hspace{.4pt}352464}}


\bibitem{HorizonWorlds}
Meta.
\newblock Horizon worlds, 2022.
\newblock \url{https://www.meta.com/horizon-worlds/}.

\bibitem{hololensWeb}
Microsoft.
\newblock Hololens, 2015.
\newblock \url{https://www.microsoft.com/en-us/hololens}.

\bibitem{hololensMRTK}
Microsoft.
\newblock Mixed reality toolkit 2, 2018.
\newblock \url{https://docs.microsoft.com/en-us/windows/mixed-reality/mrtk-unity/mrtk2/}.

\bibitem{AltspaceVR}
Microsoft.
\newblock Altspacevr, 2022.
\newblock \url{https://altvr.com/}.

\bibitem{MozillaHubs}
Mozilla.
\newblock Mozilla hubs, 2022.
\newblock \url{https://hubs.mozilla.com/}.

\bibitem{Nagendran2015}
A.~Nagendran, A.~Steed, B.~Kelly, and Y.~Pan.
\newblock Symmetric telepresence using robotic humanoid surrogates.
\newblock {\em Computer Animation and Virtual Worlds}, 26(3-4):271--280, 2015.

\bibitem{Nittala2015}
A.~S. Nittala, N.~Li, S.~Cartwright, K.~Takashima, E.~Sharlin, and M.~C. Sousa.
\newblock Planwell: Spatial user interface for collaborative petroleum well-planning.
\newblock In {\em SIGGRAPH Asia 2015 Mobile Graphics and Interactive Applications}, SA '15. Association for Computing Machinery, New York, NY, USA, 2015. doi: {{%
10\hspace{.1pt}\discretionary{.}{%
}{.}\hspace{.4pt}1145\discretionary{/}{%
}{/}2818427\hspace{.1pt}\discretionary{.}{%
}{.}\hspace{.4pt}2818443}}


\bibitem{Oyekoya2013}
O.~Oyekoya, R.~Stone, W.~Steptoe, L.~Alkurdi, S.~Klare, A.~Peer, T.~Weyrich, B.~Cohen, F.~Tecchia, and A.~Steed.
\newblock Supporting interoperability and presence awareness in collaborative mixed reality environments.
\newblock In {\em Proceedings of the 19th ACM Symposium on Virtual Reality Software and Technology}, VRST '13, p. 165–174. Association for Computing Machinery, New York, NY, USA, 2013. doi: {{%
10\hspace{.1pt}\discretionary{.}{%
}{.}\hspace{.4pt}1145\discretionary{/}{%
}{/}2503713\hspace{.1pt}\discretionary{.}{%
}{.}\hspace{.4pt}2503732}}


\bibitem{PhotonWeb}
Photon.
\newblock Photon fusion, 2019.
\newblock \url{https://www.photonengine.com/}.

\bibitem{pidel2020collaboration}
C.~Pidel and P.~Ackermann.
\newblock Collaboration in virtual and augmented reality: a systematic overview.
\newblock In {\em Augmented Reality, Virtual Reality, and Computer Graphics: 7th International Conference, AVR 2020, Lecce, Italy, September 7--10, 2020, Proceedings, Part I 7}, pp. 141--156. Springer, 2020.

\bibitem{Piumsomboon2017}
T.~Piumsomboon, Y.~Lee, G.~Lee, and M.~Billinghurst.
\newblock Covar: A collaborative virtual and augmented reality system for remote collaboration.
\newblock In {\em SIGGRAPH Asia 2017 Emerging Technologies}, SA '17. Association for Computing Machinery, New York, NY, USA, 2017. doi: {{%
10\hspace{.1pt}\discretionary{.}{%
}{.}\hspace{.4pt}1145\discretionary{/}{%
}{/}3132818\hspace{.1pt}\discretionary{.}{%
}{.}\hspace{.4pt}3132822}}


\bibitem{metaverse}
A.~Robertson and J.~Peters.
\newblock What is the metaverse, and do i have to care?, 2021.
\newblock \url{https://www.theverge.com/22701104/metaverse-explained-fortnite-roblox-facebook-horizon}.

\bibitem{RecRoom}
R.~Room.
\newblock Rec room, 2022.
\newblock \url{https://recroom.com/}.

\bibitem{Schafer2022}
A.~Sch\"{a}fer, G.~Reis, and D.~Stricker.
\newblock A survey on synchronous augmented, virtual and mixed reality remote collaboration systems.
\newblock {\em ACM Comput. Surv.}, apr 2022.
\newblock Just Accepted. doi: {{%
10\hspace{.1pt}\discretionary{.}{%
}{.}\hspace{.4pt}1145\discretionary{/}{%
}{/}3533376}}


\bibitem{stauffert2020latency}
J.-P. Stauffert, F.~Niebling, and M.~E. Latoschik.
\newblock Latency and cybersickness: Impact, causes, and measures. a review.
\newblock {\em Frontiers in Virtual Reality}, 1:582204, 2020.

\bibitem{Tanaya2017}
M.~Tanaya, K.~Yang, T.~Christensen, S.~Li, M.~O'Keefe, J.~Fridley, and K.~Sung.
\newblock A framework for analyzing ar/vr collaborations: An initial result.
\newblock In {\em 2017 IEEE International Conference on Computational Intelligence and Virtual Environments for Measurement Systems and Applications (CIVEMSA)}, pp. 111--116, 2017. doi: {{%
10\hspace{.1pt}\discretionary{.}{%
}{.}\hspace{.4pt}1109\discretionary{/}{%
}{/}CIVEMSA\hspace{.1pt}\discretionary{.}{%
}{.}\hspace{.4pt}2017\hspace{.1pt}\discretionary{.}{%
}{.}\hspace{.4pt}7995311}}


\bibitem{WiresharkWeb}
T.~W. team.
\newblock Wireshark, 1998.
\newblock \url{https://www.wireshark.org/}.

\bibitem{Teo2019}
T.~Teo, G.~A.~Lee, M.~Billinghurst, and M.~Adcock.
\newblock 360drops: Mixed reality remote collaboration using 360 panoramas within the 3d scene*.
\newblock In {\em SIGGRAPH Asia 2019 Emerging Technologies}, SA '19, p. 1–2. Association for Computing Machinery, New York, NY, USA, 2019. doi: {{%
10\hspace{.1pt}\discretionary{.}{%
}{.}\hspace{.4pt}1145\discretionary{/}{%
}{/}3355049\hspace{.1pt}\discretionary{.}{%
}{.}\hspace{.4pt}3360517}}


\bibitem{Teo2020}
T.~Teo, M.~Norman, G.~Lee, M.~Billinghurst, and M.~Adcock.
\newblock Exploring interaction techniques for 360 panoramas inside a 3d reconstructed scene for mixed reality remote collaboration.
\newblock {\em Journal on Multimodal User Interfaces}, 14, 07 2020. doi: {{%
10\hspace{.1pt}\discretionary{.}{%
}{.}\hspace{.4pt}1007\discretionary{/}{%
}{/}s12193\discretionary{%
}{-}{-}020\discretionary{%
}{-}{-}00343\discretionary{%
}{-}{-}x}}


\bibitem{Tian2023}
H.~Tian, G.~A. Lee, H.~Bai, and M.~Billinghurst.
\newblock Using virtual replicas to improve mixed reality remote collaboration.
\newblock {\em IEEE Transactions on Visualization and Computer Graphics}, 29(5):2785--2795, 2023. doi: {{%
10\hspace{.1pt}\discretionary{.}{%
}{.}\hspace{.4pt}1109\discretionary{/}{%
}{/}TVCG\hspace{.1pt}\discretionary{.}{%
}{.}\hspace{.4pt}2023\hspace{.1pt}\discretionary{.}{%
}{.}\hspace{.4pt}3247113}}


\bibitem{NetcodeWeb}
Unity.
\newblock Netcode for gameobjects, 2022.
\newblock \url{https://docs-multiplayer.unity3d.com/}.

\bibitem{RelayWeb}
Unity.
\newblock Unity relay, 2022.
\newblock \url{https://unity.com/products/relay}.

\bibitem{VRChat}
VRChat.
\newblock Vrchat, 2022.
\newblock \url{https://hello.vrchat.com/}.

\bibitem{Wang2021}
P.~Wang, X.~Bai, M.~Billinghurst, S.~Zhang, X.~Zhang, S.~Wang, W.~He, Y.~Yan, and H.~Ji.
\newblock Ar/mr remote collaboration on physical tasks: A review.
\newblock {\em Robotics and Computer-Integrated Manufacturing}, 72:102071, 2021. doi: {{%
10\hspace{.1pt}\discretionary{.}{%
}{.}\hspace{.4pt}1016\discretionary{/}{%
}{/}j\hspace{.1pt}\discretionary{.}{%
}{.}\hspace{.4pt}rcim\hspace{.1pt}\discretionary{.}{%
}{.}\hspace{.4pt}2020\hspace{.1pt}\discretionary{.}{%
}{.}\hspace{.4pt}102071}}


\bibitem{Wang2021DistanciAR}
Z.~Wang, C.~Nguyen, P.~Asente, and J.~Dorsey.
\newblock Distanciar: Authoring site-specific augmented reality experiences for remote environments.
\newblock In {\em Proceedings of the 2021 CHI Conference on Human Factors in Computing Systems}, CHI '21. Association for Computing Machinery, New York, NY, USA, 2021. doi: {{%
10\hspace{.1pt}\discretionary{.}{%
}{.}\hspace{.4pt}1145\discretionary{/}{%
}{/}3411764\hspace{.1pt}\discretionary{.}{%
}{.}\hspace{.4pt}3445552}}


\bibitem{Zhang2023}
L.~Zhang, A.~Agrawal, S.~Oney, and A.~Guo.
\newblock Vrgit: A version control system for collaborative content creation in virtual reality.
\newblock In {\em Proceedings of the 2023 CHI Conference on Human Factors in Computing Systems}, CHI '23. Association for Computing Machinery, New York, NY, USA, 2023. doi: {{%
10\hspace{.1pt}\discretionary{.}{%
}{.}\hspace{.4pt}1145\discretionary{/}{%
}{/}3544548\hspace{.1pt}\discretionary{.}{%
}{.}\hspace{.4pt}3581136}}


\bibitem{zhura2023neuroswarm}
I.~Zhura, D.~Davletshin, N.~D.~W. Mudalige, A.~Fedoseev, R.~Peter, and D.~Tsetserukou.
\newblock Neuroswarm: Multi-agent neural 3d scene reconstruction and segmentation with uav for optimal navigation of quadruped robot.
\newblock {\em arXiv preprint arXiv:2308.01725}, 2023.

\end{thebibliography}
\end{document}